# Stimulus - response curves of a neuronal model for noisy subthreshold oscillations and related spike generation

Martin Tobias Huber[1] And Hans Albert Braun[2]

[1]*Department of Psychiatry and Psychotherapy, University of Marburg, Rudolf-Bultmannstraße 8, D-35033 Marburg, Germany and* [2]*Institute of Normal and Pathological Physiology, University of Marburg, Deutschhausstraße 1, D-35033 Marburg, Germany*

Marburg, 22.09.2005

We investigate the stimulus-dependent tuning properties of a noisy ionic conductance model for intrinsic subthreshold oscillations in membrane potential and associated spike generation. On depolarization by an applied current, the model exhibits subthreshold oscillatory activity with occasional spike generation when oscillations reach the spike threshold. We consider how the amount of applied current, the noise intensity, variation of maximum conductance values and scaling to different temperature ranges alter the responses of the model with respect to voltage traces, interspike intervals and their statistics and the mean spike frequency curves. We demonstrate that subthreshold oscillatory neurons in the presence of noise can sensitively and also selectively be tuned by stimulus-dependent variation of model parameters.

## I. INTRODUCTION

Many neurons in the central and peripheral nervous system exhibit a characteristic electrical behavior on depolarising stimuli. This is characterized by intrinsic subthreshold membrane potential oscillations with related action potential generation when the oscillations reach the spike threshold [1-11]. Such oscillations are an intrinsic property of the neurons and result from the interplay of different ionic conductances [3-6]. In addition, stochastic fluctuations, which are naturally present due to the inherent noisiness of neurons, play an important role for the response behavior. When oscillations operate close to the spike threshold, the noise can essentially determine whether a spike is triggered or not and even little stochastic fluctuations can initiate action potential generation [8,12-18].

The role of subthreshold oscillations for neuronal function is different depending on location and function of the involved neuron. That is, a peripheral neuron or sensory receptor responsible for sensing environmental sensory stimuli can use stimulus-dependent modulations of subthreshold oscillations for that purpose [8]. Probably the most interesting example for such encoding with oscillations and noise are shark electroreceptors. These receptors can encode electrical and thermal stimuli and it was demonstrated experimentally that these electroreceptors do use intrinsic subthreshold membrane potential oscillations in cooperation with stochastic fluctuations to improve their encoding sensitivity to these stimuli. Moreover, selective stimulus-dependent modulation of the spiking probability per oscillation cycle and oscillation frequency is used for the differential encoding of the two sensory modalities [8]. This is because electrical stimuli rather selectively modulate the spiking probability whereas temperature alters both the spiking probability and the frequency. Such temperature-dependent effects on neuronal oscillations are also known from temperature-encoding by peripheral themoreceptors [12,13,16,19,20].

Neurons located in central structures such as the entorhinal cortex [6] or amygdala [4,5] are supposed to use the oscillations for the timing of collective rhythmic behaviours e.g. the limbic theta rhythm and the generation of synchronized responses [2,3,21-24]. Fine tuning of oscillatory responses by neuromodulatory substances e.g. due to cholinergic and dopaminergic alteration of oscillatory dynamics were described [25,26] and could also be used for proper adjustment of signal encoding properties in the central nervous system. The importance of noise for subthreshold oscillations was considered also with regard to central neurons [18,27].

However, subthreshold oscillations can also become important under pathological conditions. This is the case with pain-sensitive (nociceptive) dorsal root ganglion neurons. In such nociceptive neurons, previous nerve injury enhances subthreshold oscillatory activity and increases the number of subthreshold oscillatory cells. These subthreshold oscillations seem necessary for sustained spiking and ectopic spike discharge and hence development of neuropathic pain states [7,9,10,28-32]. Interestingly, spike patterns and interspike interval histograms (ISIHs) recorded from such injured sensory neurons [10] resemble the ones recorded form shark electroreceptors and subthreshold oscillating thermoreceptors, again indicating a significant influence of stochastic fluctuations on oscillatory responses [8]. The productive role of stochastic fluctuations and specifically stochastic resonance for sensory encoding and biological systems in general has attracted much attention since the first paper on stochastic resonance in a sensory receptor was published by Moss and coworkers [33],. A multitude of papers than has been demonstrating productive qualitative noise effects in experimental and theoretical studies reaching from the molecular, cellular, systems and even to the behavioural level and disease models [8,12,18,33-39,40-55].

The effect of tuning of the subthreshold oscillations by alteration of ionic conductance parameters and the resulting effects on signal encoding and neuromodulation have attracted less interest, perhaps with the exception of impulse pattern modifications in peripheral thermoreceptors [12,16,19]. Therefore, in the present paper we concentrate on response behaviors and tuning curves resulting from modulation of neuronal oscillations in the presence of a fixed noise level. We use a computational approach and consider the modulatory features of a minimal yet physiologically plausible ionic conductance model for noisy subthreshold oscillations and associated spiking which is able to reproduce some of the essential encoding and modulatory properties observed experimentally. The paper is organized as follows. We start with the situation where depolarising electrical current application leads to noisy subthreshold oscillations and action potentials in our model neuron and demonstrate the characteristic voltage traces, interspike interval plots and interspike interval histograms. In the following sections we than investigate systematically how variation of the applied current $I_{app}$, the noise intensity D, the subthreshold potassium conductance $g_{Ks}$ and the temperature T change the model neuron´s responses with respect to voltage traces, ISIs, ISIHs and corresponding mean spike frequency curves. We end with a short summary and concluding remarks.

## II. MODEL

The model is a modification of a previous ionic conductance model for subthreshold oscillations and action potential generation [39,56]. The model consists of two sets of simplified sodium and potassium conductances operating at two different membrane potentials and two different time scales: a Hodgkin/Huxley-type spike encoder is represented by rapid, high-voltage activating $g_{Na}$ and $g_K$, whereas the subthreshold oscillations essentially depend on the interplay of a persistent sodium conductance, $g_{Nap}$, and a subthreshold potassium conductance, $g_{Ks}$.

$$C_M \, dV/dt = -I_l - I_{Nap} - I_{Ks} - I_{Na} - I_K + I_{app} + \zeta \tag{1}$$

with $C_M = 1 \mu F/cm^2$ the membrane capacity, V the membrane voltage, $I_l = g_l (V-V_l)$ a leak current with $g_l = 0.1 \, mS/cm^2$ and $V_l = -60 \, mV$. $I_{app}$ is injected current and $\zeta$ is gaussian white noise with zero mean and dimensionless intensity D given as $\zeta = (-4 \, D \, dt \, \ln(a))^{1/2} \cos(2\pi b)$ [58] where a and b (b in mV) are random numbers (0 to 1) and dt is the integration time step.

The voltage-dependent currents $I_{Nap}$, $I_{Ks}$, $I_{Na}$ and $I_K$ are modeled as

$$I_i = \rho \, g_i \, a_i \, (V - V_i) \tag{2}$$

with $\rho$ a temperature-like scaling factor, $g_i$ the respective maximum conductances (i denotes Na, Nap, K, Ks), $a_i$ the voltage-dependent activation variables, V the membrane potential and $V_i$ the respective Nernst potentials. The activation variables are given as

$$da_i / dt = \varphi \, (F_i - a_i)/\tau_i \tag{3}$$

with

$$F_i = 1/\{1 + \exp[-s_i \, (V - V_{0,i})]\} \tag{4}$$

where $\varphi$ is a temperature-like scaling factor, $\tau_i$ are time constants, $s_i$ the slope and $V_{0,i}$ the half-activation potentials. Activation of $I_{Na}$ is instantaneous, thus $a_{Na} = a_{Na\infty}$. The temperature-like scaling factors $\rho$ (for ionic currents) and $\varphi$ (for ionic kinetics) are given as $\varphi = 3.0^{(T-25)/10}$ and $\rho = 1.3^{(T-25)/10}$ with T the temperature in $^{o}C$ [57].

The system of equations was solved numerically by use of the forward Euler integration method with stepsize adjusted to 0.1 ms according to the implementation of Fox et al. [58] given as $V_{t+dt} = V_t + f(V) \, dt + \zeta$ where $f(V) = I_{app} + \sum - a_i \, g_i \, (V - V_i)$. Numerical parameter values: $V_{Na} = 50$, $V_K = -90$, $g_{Na} = 2.0$, $g_K = 2.0$, $g_{Nap} = 0.4$, $g_{Ks} = 2.0$, $s_{Na} = s_K = s_{Nap} = s_{Ks} = 0.25$, $\tau_K = 2.0$, $\tau_{Nap} = 10$, $\tau_{Ks} = 50$, $V_{0Na} = V_{0K} = -25$, $V_{Nap} = V_{0Ks} = -40$. Systems of units is ms, mV, mS/cm$^2$, mA/cm$^2$.

## III. RESULTS

Our starting point is the model depolarized by an applied current $I_{app}$ which results in subthreshold oscillations of the membrane potential (fig.1). In this situation, noise becomes important for spike generation because it essentially determines whether a spike is triggered or not. Without noise the model would remain completely subthreshold in the simulations shown in figure 1. With noise, we observe a characteristic response pattern characterized by subthreshold oscillations and oscillations which occasionally trigger action potentials. In the plot of successive interspike intervals this behavior is reflected by separated bands of intervals (fig. 1b). The interspike interval histogram shows the well-known multimodal interval distribution with interval peaks located at approximate integer multiples of a basic oscillation period (fig. 1c).

The amplitude of the oscillations and hence the likelihood for spike generation depends on the level of depolarizing current $I_{app}$ (fig. 2). At low levels of $I_{app}$ only rarely a spike is triggered whereas on higher levels of $I_{app}$ an oscillation only occasionally fails to trigger a spike (fig. 2a). Accordingly, in the ISIHs, the first interval peak representing the basic oscillation period increases with increasing depolarization and the multimodality decreases (fig. 2b). The continuous ramp-like change of $I_{app}$ and its effect on the oscillatory responses is demonstrated on hand of the time plot of interspike intervals (fig. 2c). First, intervals are scattered to long durations and are more or less triggered randomly by the noise. In intermediate ranges, distinctive bands of intervals occur resulting from the depolarization induced oscillations and associated periodic spike generation. Finally, when almost all oscillations reach spike threshold, the intervals become concentrated on the band which represents the basic oscillation period.

In the whole range from occasional spikes to fully periodic spiking the oscillation frequency is not much affected. This is indicated by an almost unchanged location of the interspike interval peaks and bands, respectively (fig. 2b,c). The reason is that the model operates close to the spike threshold where small changes in membrane depolarisation have significant effects on the probability of spike generation without similar pronounced effects on the oscillation frequency. This effect is also reflected in the averaged measure of the mean spike frequency curve. In the mean spike frequency (F) versus applied current ($I_{app}$) plot, this is the range where the increase in F is most pronounced (fig. 3a) and where under fully deterministic conditions (D = 0; dashed curve in fig.3a), F jumps in a steplike way when $I_{app}$ crosses a threshold value ($I_{app} > 1.34$ mA/cm$^2$ in our model). The noise smoothes the step-like nonlinearity of the deterministic curve. Increasing the noise intensity reduces the slope and in this way linearizes the relation. The response range increases in that way, however, on cost of the response sensitivity. Figure 3a also shows that increasing depolarizing current monotonically increases the spike frequency also in the suprathreshold response range. The increase of F in the suprathreshold range now results from an $I_{app}$-induced acceleration of the oscillation frequency and not from modulation of the spiking probability as it is the case in the range close to the spike threshold.

The effect of the noise intensity on the mean spiking frequency for different fixed values of $I_{app}$ is shown in figure 3b. Two principally different behaviors are observed depending on whether the model is in a subthreshold or suprathreshold state. In the subthreshold range, F rises monotonically with the noise intensity. Here the noise can initiate spike generation in otherwise subthreshold oscillations and F rises as higher noise intensities lead to a higher spiking probability. In the suprathreshold range, the situation is different as F initially decreases with increasing noise intensity and increases again with higher levels of noise. In the suprathreshold situation, some noise can suppress spiking in former spike-triggering oscillations. However, with high noise levels, the excitatory effect of the noise becomes dominating and accordingly F rises monotonically. In both cases, the initial increase or decrease of the mean frequency is most pronounced at low to moderate noise intensities where the noise interacts with the oscillatory dynamics (fig. 3b). High noise intensities then overwhelm the dynamics, destroy the oscillations and spiking is no longer related to an underlying periodic process.

We next consider the modulatory effect of ionic conductances underlying the subthreshold oscillations. As example we chose the subthreshold potassium conductance $g_{Ks}$ and vary the maximum conductance $g_{Ksmax}$. Experimentally, this would correspond to the application of potassium channel blockers like 4-aminopyridine or, physiologically, to the action of channel-blocking neuromodulatory substances. Figure 4 shows the change of interspike intervals and the mean spike frequency in response to a ramp-like change of $g_{Ksmax}$ (fig. 4c,d) together with voltage traces and ISIHs at three different values of $g_{Ksmax}$ (fig. 4a,b). A reduction of the $g_{Ksmax}$ has a depolarizing effect on the membrane voltage. In turn, subthreshold oscillations develop ($g_{Ksmax} = 2.1$ mS/cm$^2$) leading to the characteristic multimodality of the ISIHs and interspike interval plots. With further decreasing $g_{Ksmax}$ the subthreshold oscillations rise in amplitude until finally periodic spiking occurs. Correspondingly, the ISIHs and interpike plots change to unimodal distributions ($g_{Ksmax} = 2.0$ -> $1.9$ mS/cm$^2$).

The mean spiking frequency F increases monotonically and approximately sigmoidal with decreasing $g_{Ksmax}$ values (fig.4d, fig. 5a). The response range than can be tuned to lower or higher $g_{Ksmax}$ values by respective adjustment of the amount of $I_{app}$, that is by adjusting the preactivation level with $I_{app}$ (fig. 5a). Higher values of $I_{app}$ shift the F curve to higher $g_{Ksmax}$ values and vice versa. Similarly, the mean spiking frequency in dependence to $I_{app}$ can be tuned to lower or higher values by adjustment of the $g_{Ksmax}$, that is by adjusting the preactivation level with the $g_{Ks}$ (fig. 5b). Reducing the $g_{Ksmax}$ decreases the total amount of repolarizing ionic current and for this reason less applied current is needed for membrane depolarization. Accordingly, the F curve is shifted to lower $I_{app}$ values. Increasing the $g_{Ksmax}$ has the opposite effect because it increases the repolarizing ionic current and therefore the amount of $I_{app}$ needed for depolarisation.

The last part of the results considers how temperature scaling of the ionic conductances alters the responses of the model. Voltage traces and ISIHs at three different steady temperatures are shown in figure 6a,b. A time plot of interspike intervals in response to a ramp-like temperature change and the corresponding mean spiking frequency is demonstrated in figure 6b. Temperature scaling alters predominantly the time constants $\tau_i$ of the ionic currents and to a minor degree the values of the maximum ionic currents (see eqs. 2 and 3 in Sec. II). The temperature effect on the activation time constants in turn has a pronounced influence on the frequency of the subthreshold oscillations and action potentials (fig. 6a). The effects on amplitudes of oscillations and action potentials are to some extent compensated by the changes of the maximum ionic currents which decrease with decreasing T and vice versa. The dominant effect on the oscillation frequency leads to characteristic changes in the ISIHs and interspike interval plot (fig. 6b,c). The acceleration of the oscillation frequency with rising temperature increases the number of subthreshold oscillations. This is because oscillations then become too fast to initiate action potentials during an oscillation cycle (fig. 6a,b). In the ISI time plot, the bands of intervals successively change to shorter interval durations and with high temperatures a scattering to longer interval durations occurs. The latter results from the accelerating oscillations which more and more fail to trigger spikes (fig. 6c).

The acceleration of the oscillation frequency with associated changes in the spiking probability has a characteristic effect on the mean frequency curve which is very different to the stimulus - response relations obtained by variation of $I_{app}$ or $g_{Ksmax}$. In the case of temperature variation, F(T) first rises with increasing temperature but than passes through a maximum and declines again on further increasing T (fig. 6d and fig. 7a). The rise in F is due to the increase of the oscillation frequency and which in turn results from temperature-accelerated ionic kinetics. The following decrease of F results from the increasing number of faster and faster subthreshold oscillations which successively fail to trigger spikes. The maximum value of F as well as the slope and response range of the F(T) curve also depend on the level of depolarizing current $I_{app}$ (fig. 7a). When the model neuron is in a more depolarized state, oscillatory activity is more pronounced and suprathreshold. In this case, oscillations can become much faster (and F much higher) until they fail to trigger spikes when becoming subthreshold (e.g. $I_{app}$ = 2.5 mA/cm$^2$ versus $I_{app}$ = 1.3 mA/cm$^2$). It should be noted that noise enlarges the response range with respect to T as noise can induce spiking in subthreshold oscillations. Without noise, F(T) would rise monotonically up to a certain limit and than would immediately drop to zero when oscillations become subthreshold.

The corresponding F versus $I_{app}$ curves at different fixed temperature levels also reflect the temperature-dependent variation of the oscillation frequency and associated spiking probability per oscillation cycle (figure 7b). The F-$I_{app}$ curves increase monotonically with increasing $I_{app}$. The obtainable maximum frequencies depend on the applied temperature range and higher F values are achievable with higher temperatures. However, high temperature levels need higher levels of depolarization (larger $I_{app}$) because of the very short suprathreshold periods of the fast oscillations (fig. 7b: T = 35 $^{o}$C). Further increases of $I_{app}$ in the high temperature range than result in a steep increase of the mean spike frequency and which is much steeper than the ones obtained at lower temperature ranges. These results demonstrate that temperature-dependent modulation of noisy oscillations (i.e. oscillation frequency and spiking probability) has pronounced effects on temporal response patterns (voltage traces, ISIs, ISIHs) as well as on the mean spike frequency curve and its respective slope (or gain).

## IV SUMMARY AND CONCLUSIONS

Our results presented here demonstrate that a minimal ionic conductance model for noisy subthreshold oscillations and related spike generation exhibits interesting modulatory features on physiologically relevant parameter variations. Modulatory effects thereby occur in the temporal response patterns (voltage traces and ISI distributions) as well as when the mean spike frequency is considered as a measure for signal or stimulus encoding. All of the described effects were obtained by modulation of *deterministic* parameter values (i.e. $I_{app}$, $g_{Ksmax}$ or the temperature scaling coefficients) which resulted in variation of the oscillation frequencies and amplitudes and associated changes in the spiking probability per oscillation cycle.

The full range of responses thereby is only possible because the model includes noise. This is because only the modulation of noise-induced transitions from subthreshold oscillations to spike-triggering oscillations allows for the pronounced changes in responses on even small parameter changes demonstrated by the changes in spike frequency and ISI statistics. However, although noise is an essential ingredient, the observed modulatory effects resulting from modulation of applied current, ionic conductance levels or temperature scaling are not critically dependent on the actual noise intensity. That is, similar tuning behaviors are obtained at very different noise intensities indicating the robustness of the observed behaviors. In addition to this, a variety of modulatory effects can be expected from tuning of the noise intensity such as linearization of responses and stochastic or coherence resonance effects pointing to even more potential modulatory capabilities resulting from the interplay of oscillations and noise (see e.g. refs. [59-62]).

Many studies have addressed the importance of noise and in particular noise tuning in neurons and neuronal models and also some studies have addressed deterministic subthreshold oscillations [3,18,24]. Our study shows that tuning of the oscillations themselves or more correct the deterministic parameters underlying the oscillatory responses, plays an equally important role for the overall encoding and modulatory properties. In addition, the effects demonstrated for a simple ionic conductance model have their real counterparts in biology. The deterministic parameters have a clear physiological meaning such as e.g. neuromodulation of a potassium conductance in a CNS neuron [25,63], encoding of environmental electrical or thermal stimuli by a sensory receptor neuron [8] or the pathophysiological relevant oscillatory response of DRG cells to nerve injury resulting from changed ionic conductance balances [7,9,29].

So far, DRG cells are the best example for a pathophysiological relevance of noisy subthreshold oscillations and shark electroreceptors are the best example for direct and differential encoding of sensory stimuli using this mechanism. In both cases, the interplay of oscillations and noise provide for a large sensitivity and in case of the electroreceptor also for differential encoding of electrical and thermal sensory stimuli by selective modulation of the noisy oscillations. Although several differences exist between cortical neurons, DRG cells and sensory receptors including the respective specific biological equipment, our study shows that some of the relevant behaviors can be represented with a simplified and generalized ionic conductance model. Being inherently noisy and tentatively oscillatory due to their nonlinearity it seems that neurons have learned to use the two features for neuromodulation and signal encoding under physiological and pathophysiological conditions. Our study here is limited to neuromodulation and stimulus encoding at the single neuron level. We suggest that future studies should also consider the described effects with respect to coupled neurons and larger-scale neuronal networks.

**Acknowledgment:** We acknowledge financial support by the EU Network of Excellence NoE 005137 BioSim (RA5WP10).

**Figure 1**

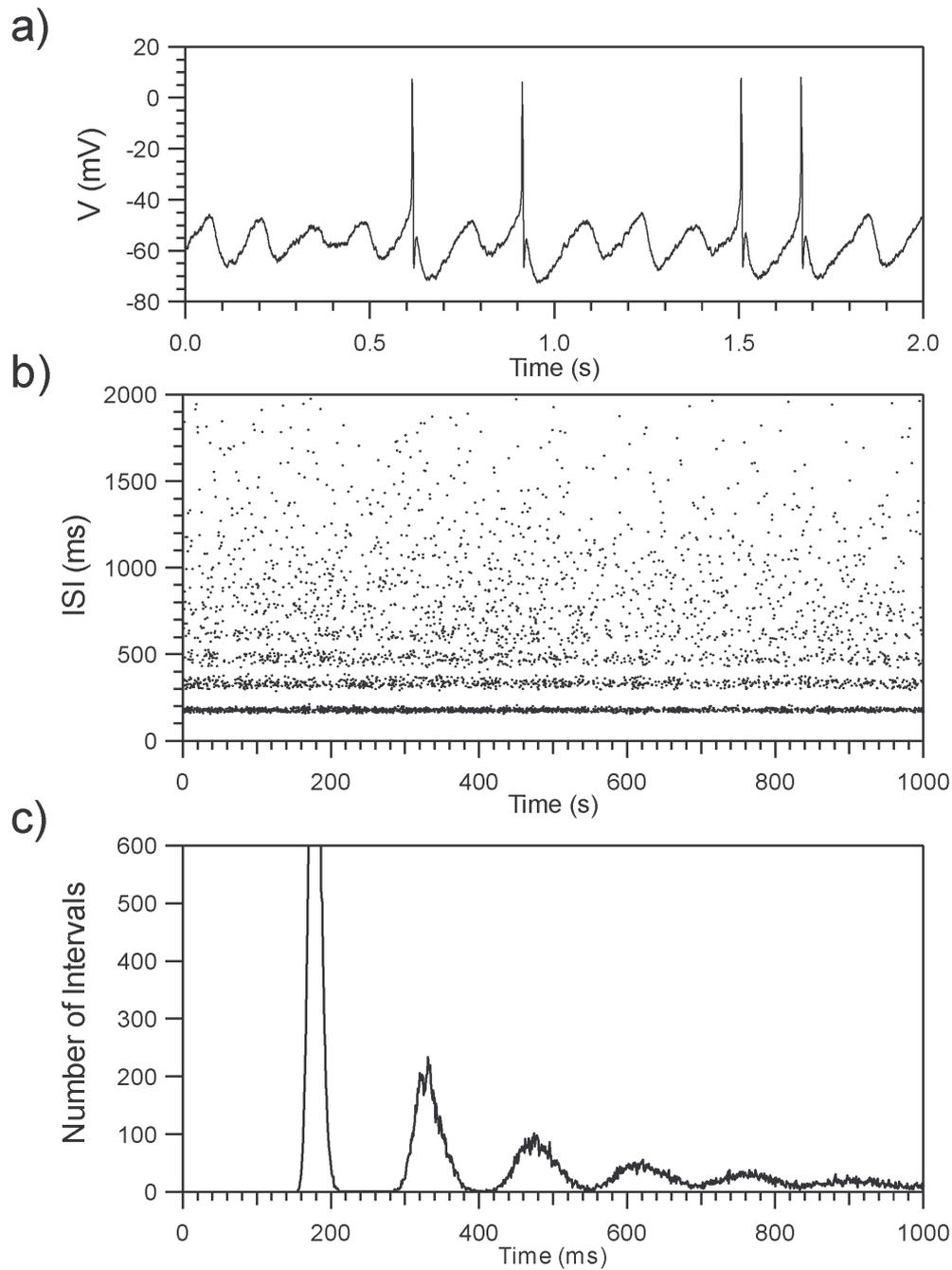

Figure 1: Noisy subthreshold oscillations and related spike generation generated by the model. a) time sequence of the membrane voltage b) time plot of successive interspike intervals from a longer simulation run and c) the corresponding interspike interval histogram. Simulation time t = 1000 s, $I_{app}$ = 1.3 mA/cm$^2$, noise intensity D = 0.1.

**Figure 2**

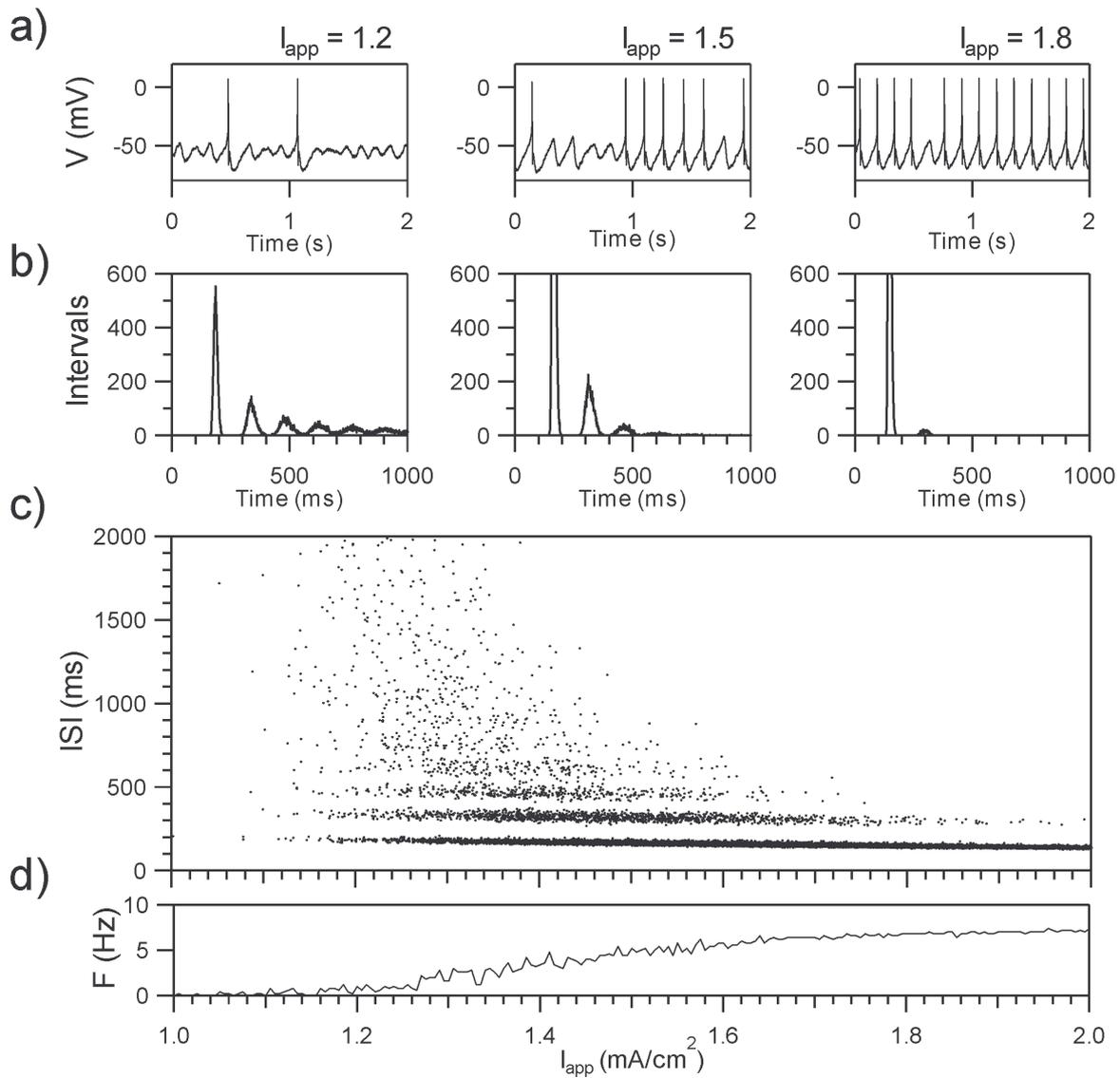

Figure 2: Effect of depolarizing applied current $I_{app}$. a) voltage traces and b) interspike interval histograms (ISIHs) for $I_{app}$ = 1.2, 1.5 and 1.8 mA/cm$^2$, c) time plot of successive interspike intervals (ISIs) and d) mean spike frequency (F) on response to a ramp-shaped change of the applied current (time = 1000 s, increment $\Delta I_{app}$ = 0.001 mAcm$^{-2}$ms$^{-1}$, noise intensity D = 0.1).

**Figure 3**

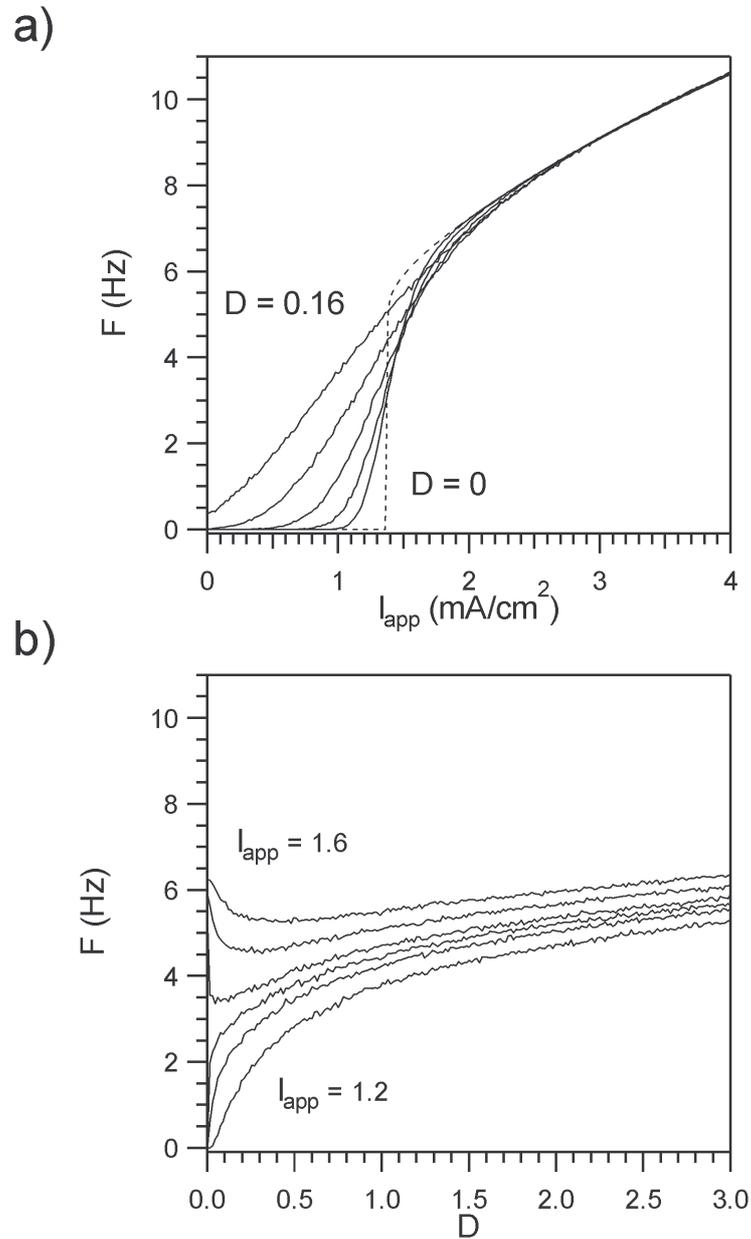

Figure 3: a) Mean spike frequency F (Hz) versus $I_{app}$ (mA/cm$^2$) for different values of the noise intensity (D = 0, 0.1, 0.2, 0.4, 0.8, 0.16), b) mean spike frequency F versus D for different values of the applied current ($I_{app}$ = 1.2, 1.3, 1.35, 1.4, 1.5, 1.6 mA/cm$^2$).

**Figure 4**

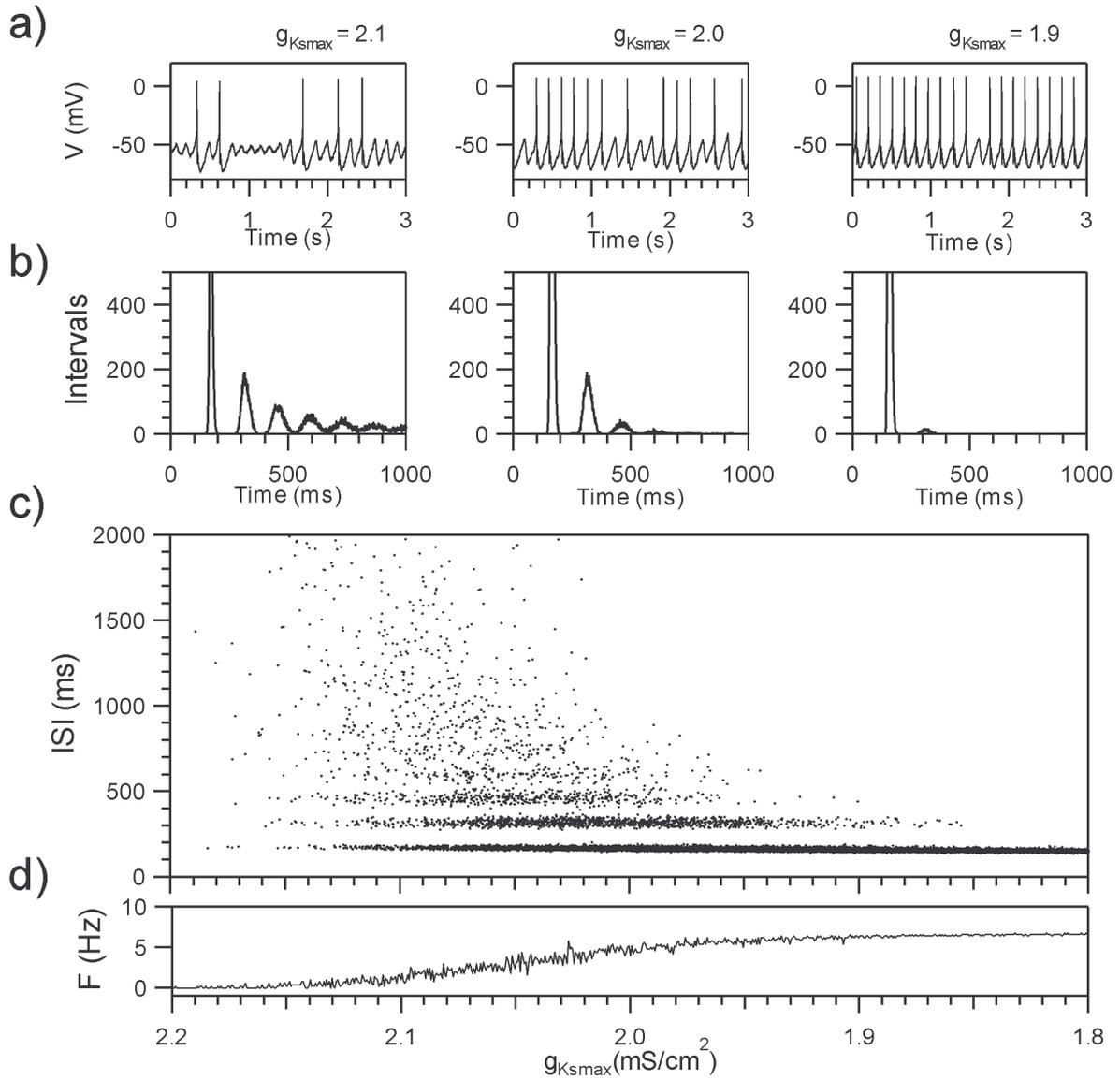

Figure 4: Effect of potassium conductance $g_{Ks}$. a) voltage traces and b) interspike interval histograms (ISIHs) for $g_{Ksmax}$ = 2.1, 2.0 and 1.9 mS/cm$^2$, c) time plot of successive interspike intervals (ISIs) and d) mean spike frequency (F) on response to a ramp-shaped change of $g_{Ksmax}$ (time = 1000 s, increment $\Delta g_{Ksmax}$ = 0.0004 mS/cm$^{-2}$ms$^{-1}$, noise intensity D = 0.1).

**Figure 5**

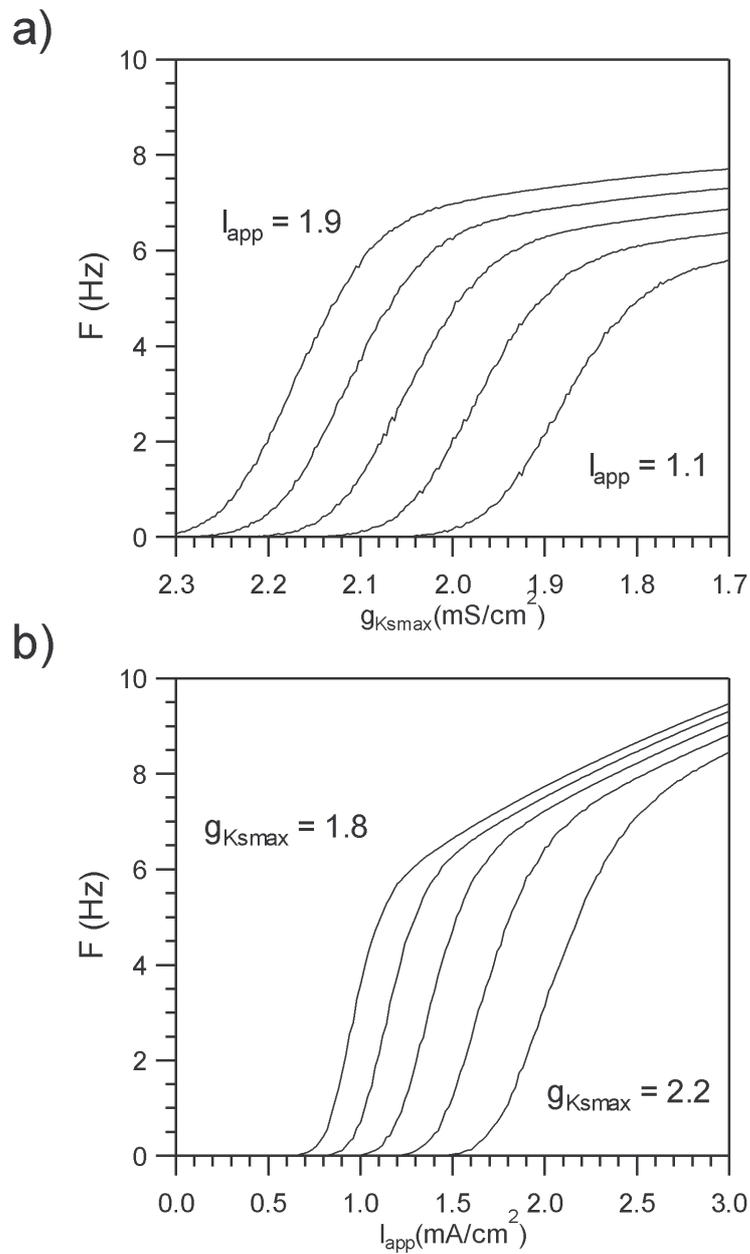

Figure 5: a) Mean spike frequency F (Hz) versus $g_{Ksmax}$ (mS/cm$^2$) for different values of applied current ($I_{app}$ = 1.1, 1.3, 1.5, 1.7, 1.9 mA/cm$^2$), b) mean spike frequency F (Hz) versus $I_{app}$ (mA/cm$^2$) for different values of $g_{Ksmax}$ (1.8, 1.9, 2.0, 2.1, 2.2 mS/cm$^2$).

**Figure 6**

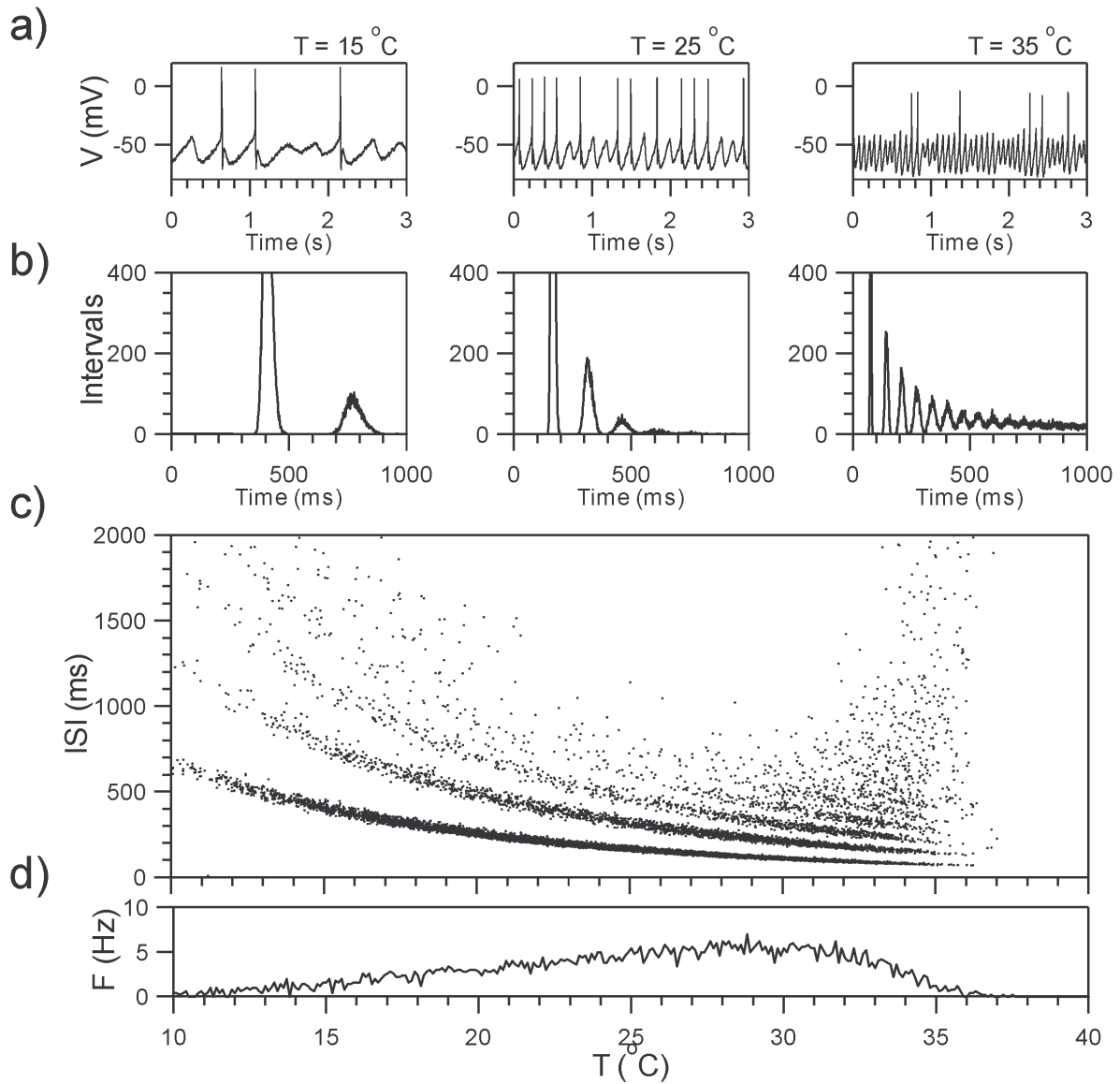

Figure 6: Effect of temperature. a) voltage traces and b) interspike interval histograms (ISIHs) for T = 15, 25, 35 °C, c) time plot of successive interspike intervals (ISIs) and d) mean spike frequency (F) on response to a ramp-shaped change of T (time = 1000 s, increment ΔT = 0.03 °C/ms, noise intensity D = 0.1).

**Figure 7**

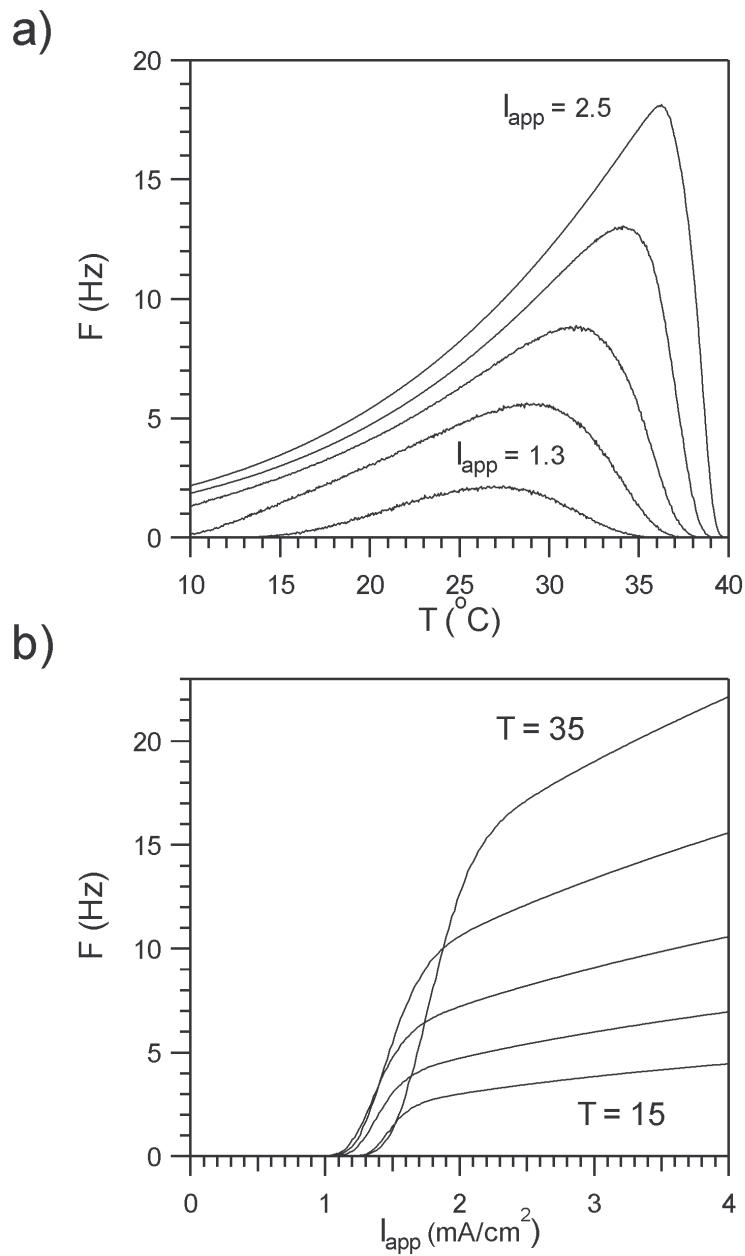

Figure 7: a) Mean spike frequency F (Hz) versus temperature T (°C) for different values of applied current ($I_{app}$ = 1.3, 1.5, 1.7, 2.0, 2.5 mA/cm$^2$), b) mean spike frequency F (Hz) versus $I_{app}$ (mA/cm$^2$) for different values of the temperature (T = 15, 20, 25, 30, 35 °C).